# Virtual synchronous generator of PV generation without energy storage for frequency support in autonomous microgrid

Cheng Zhong[a], Huayi Li[a], Yang Zhou[a], Yueming Lv[a], Jikai Chen[a], Yang Li[a]

[a] Key Laboratory of Modern Power System Simulation and Control & Renewable Energy Technology (Ministry of Education), Northeast Electric Power University, Jilin ,132012, China

**Abstract**: In autonomous microgrids frequency regulation (FR) is a critical issue, especially with a high level of penetration of the photovoltaic (PV) generation. In this study, a novel virtual synchronous generator (VSG) control for PV generation was introduced to provide frequency support without energy storage. PV generation reserve a part of the active power in accordance with the pre-defined power versus voltage curve. Based on the similarities of the synchronous generator power-angle characteristic curve and the PV array characteristic curve, PV voltage $V_{pv}$ can be analogized to the power angle $\delta$. An emulated governor (droop control) and the swing equation control is designed and applied to the DC-DC converter. PV voltage deviation is subsequently generated and the pre-defined power versus voltage curve is modified to provide the primary frequency and inertia support. A simulation model of an autonomous microgrid with PV, storage, and diesel generator was built. The feasibility and effectiveness of the proposed VSG strategy are examined under different operating conditions.

## 1. Introduction

Microgrids are emerging as a cost-effective solution for the integration of distributed generations (DGs) in the recent decades. However, considering the high penetration of DGs, the microgrid is an electrical system having a low inertia and a lack of FR [1]. In particular, in the autonomous mode of operation, the active power of DGs is controlled locally to maintain a balance between the generation and demand while maintaining the frequency within the predefined range. Many DG devices, such as in PV and wind turbine generation, interfaced to the grid by power electronic converters, do not have inertia and FR abilities [2-3]. Thus, FR is a critical problem for microgrids, and a large capacity for energy storage and diesel with high cost is usually required.

Droop-based control is a significant solution for microgrids because of the salient features of communication-free and plug-and-play capability [4-5]. Conventionally, active power -frequency (*P-f*) or frequency-active power (*f-P*) droop control is deployed to generate frequency support for DGs. However, droop-based DGs still lack inertia unlike the synchronous generators (SGs), which is detrimental to the dynamics of the frequency. The concept of virtual synchronous generator (VSG) control has been presented recently [6–8]. VSG mimics the behaviors of conventional SGs and adds virtually inertia to the system [9]. A comparison of dynamic characteristics between the VSG and droop control is discussed in [10]. It is suggested that VSG inherits the advantages of droop control and provides inertia support for the system. A comprehensive survey of the VSG and the existing topologies are given in [11]. An adaptive virtual inertia VSG, which combines the merits of large inertia and small inertia, is presented in [12].

However, VSG is usually applicable for DGs with a constant DC voltage and pre-known active power reserve, such as energy storage devices, or PV/wind turbine generators (WTG) [13] with energy storage. For a PV system or WTG without energy storage, the output power is random and limited by the environmental conditions. PV system has no power reserve or inherent rotor inertia. Furthermore, for the two-stage PV system, instead of the mimic swing equation control in VSG, its DC-link voltage loop is required through the AC/DC inverter. This means that the conventional VSG does not adapt to two-stage PV system. This paper focuses on designing an improved VSG control for PV systems without energy storage.

In this paper, to introduce the inertia and FR abilities for two-stage PV generation without energy storage, a novel VSG control method is proposed. This method maintains a part of the active power by PRC control and combines VSG technology to enable the PV system to support FR in the island microgrid. The salient features of the proposed VSG are as follows. (1) A pre-definition power versus voltage curve is utilized to realize PRC control. It is a simple and does not require the use of irradiance measurement method. (2) Unlike the typical VSG, which is used for a DC-AC inverter, the proposed method is applied to a DC–DC converter. (3) The governor (droop control) and the emulated swing equation control are constructed to provide frequency support.

The remainder of this paper is organized as follows. Section 2 briefly summarizes the existing PV FR strategies. Section 3 introduces the modeling of a PV diesel-storage island microgrid, which is from a real microgrid of the YongXing island microgrid located in Hainan, China. Section 4 details the proposed VSG control for the PV system. Simulation results are presented in Section 5. Finally, a brief conclusion is given in Section 6.

## 2. FR strategies for PV system

Many FR strategies for PV system have been proposed and can be categorized into two categories: installing energy storage or employing a power reserve control.

The former category, PV is combined with energy storage and the power reserve is provided from the energy storage. In [14], a novel VSG control strategy for PV-storage grid-connected system was proposed, which the energy storage unit implements the maximum power point tracking control and the photovoltaic inverter implements a virtual synchronous generator algorithm which can both provide inertial and primary frequency support for microgrid. In [15], three parallel VSG based PV systems integrated with battery storage systems are used to analyze the frequency response and its stability. This method use genetic algorithm (GA) optimization to optimize the values for VSG control parameters, which makes this method very complex. With the help of proper control philosophies implemented in a battery storage system the frequency disturbances can be handled effectively.

The latter category, FR is incorporated into PV itself by power reserve control (or called de-loading control) to enhance the operation of the stand-alone microgrid [16-18]. Numbers of power reserve control methods have been proposed for PV system. The modified Perturb and observe (P&O) maximum power point tracking (MPPT) algorithm was proposed in [19] to operate PVs at a constant power level (below MPP). To speed up the tracking process, an adaptive flexible power point tracking algorithm was proposed in [20]. In [21–22], irradiance and temperature measurements are utilized to extrapolate the mathematical PV modeling parameters to the actual conditions; however, the irradiance sensor is expensive. In [23–26], mathematical PV modeling is applied to voltage and/or current measurement. In particular, [23-24] linear-quadratic modeling is applied to two or three measurements. A single-diode PV model was employed, and a ripple control method was utilized to estimate the model parameters by the least square method in [25]. Delta power control strategy, combination of maximum power point tracking (MPPT) and constant power generation (CPG) modes was proposed in [26]. But, it is only adapted for the multistring PV inverter that the mismatch between each PV string is very small. [27] proposed an iterative estimation method that only required one sample point. However, it was compromised by the required off-line PV model parameters. Further, the inertia response is depended on the $df_g/dt$ measurement which may deteriorate the FR performance because of the filter delay and attenuation. In [28], the MPP under dynamic conditions is monitored by an artificial neural network-based estimator and the reserve is adjusted using a fuzzy reserve controller.

**Table 1. Comparison of PV FR strategies**

| Ref | Installing energy storage | How to obtain $P_{mpp}$ | Additional sensors requirement | Implantation | Inertia support | $df_g/dt$ measurement |
|---|---|---|---|---|---|---|
| [14] | Yes | No required | No | Simple | Yes | No |
| [15] | Yes | No required | No | Complex | Yes | No |
| [19] | No | Modified P&O algorithm | No | Simple | No | No |
| [20] | No | Adaptive FPPT algorithm | No | Simple | No | No |
| [21,22] | No | The off-line polynomial of $P_{mpp}$ versus irradiance and temperature | Yes | Simple | No | No |
| [23,24] | No | Quadratic PV array modeling | No | Complex | No | No |
| [25] | No | Numbers of $I_{pv}$ and $V_{pv}$ are sampled, LSQ iteration estimate. | No | Complex | No | No |
| [26] | No | Master-slave modes | No | Simple | No | No |
| [27] | No | Linear relationship between $I_{sc}$ and $P_{mpp}$ | No | Simple | Yes | Yes |
| [29] | No | Master-slave modes | No | Simple | Yes | Yes |
| [30] | No | Peer-to-peer or master-slave modes | No | Simple | Yes | Yes |
| Proposed | No | De-loaded power-voltage curve | No | Simple | Yes | No |

In fact, a few studies have attempted to apply VSG control for PV systems without energy storage [29-30]. In [29], a VSG with power reserve control (PRC) is presented. Maximum power point estimation (MPPE) is used, and a part of the active power is reserved through direct power control. An additional power based on droop characteristic (called PRC-VSG control in [29]) is generated for the frequency support. Similarly, in [30], two kinds of PRC methods for the peer-to-peer PV systems or master-slave PV systems are used to reserve a part of the power. An additional active power based on frequency deviation and rate of change of frequency (also called VSG) is added to provide frequency support. However, these previous PV-VSG strategies do not mimic the conventional behavior of the SGs, such as the swing equation.

For the convenience of analysis and observation, the main features of the previous FR strategies for PV system are summarized in Table 1.

In addition, partial shading condition is also a critical problem, especially for large scale photovoltaic power plants [31]. Partial shading conditions make the power versus voltage (P-V) curve of PVs has multiple peaks instead of a single peak. In [32], a curve fitting method is used to estimate the MPPs (local maximum point and global maximum point) and inflection points (IP) which makes this method very complex. [33] proposed a simple power reserve control under partial shading conditions that avoids the blind scanning of the power versus voltage curve and scans only across the generated reference points. In this study, the small capacity of the PV system is considered and the partial shading conditions are neglected due to the small area of PV array.

## 3. PV-diesel-storage island microgrid modeling

Fig. 1 depicts the schematic diagram of the island microgrid located in Yongxing Island, China, which includes a diesel unit (DU), three two-stage PV generations, a battery storage unit, and distributed loads. Here, the microgrid is not connected to any utility grid, and it operates independently as a stand-alone system. The capacity configuration of the microgrid is listed in Table 2.

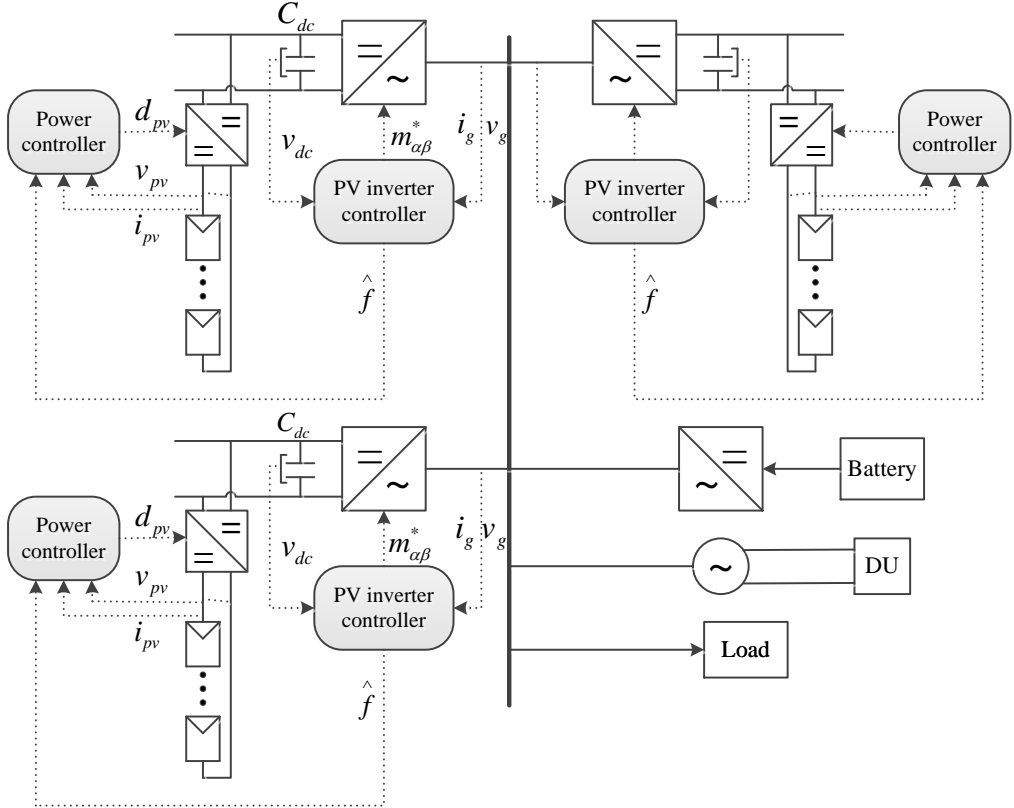

Fig. 1. Schematic diagram of the microgrid system

Moreover, the FR of the island microgrid depends on the DU and battery storage unit, whereas the PV generator operates in maximum power point tracking (MPPT) and does not participate in FR [34].

**Table 2. Microgrid configuration**

| DG | Rated output power | Rated voltage | Rated frequency |
|---|---|---|---|
| Diesel unit | 50 kW | 5 kV | 50 Hz |
| PV Generator | 15 kW | 5 kV | 50 Hz |
| Battery storage | 10 kW | 5 kV | 50 Hz |

### 3.1 Diesel unit

The simple DU model employed in this study is illustrated in Fig. 2. It is composed of a DU, valve actuator, synchronous generator and exciter. The time constants of the DU and the valve actuator are represented by $T_D$ and $T_{SM}$, respectively. The parameter of the governor is the droop constant $R$. Inputs to the synchronous generator are the mechanical power $P_D$ and the terminal voltage $v_t$, and the outputs are the generator speed $\omega_D$ and the electrical output power $P_e$. The parameter values are given in Table A1 of Appendix.

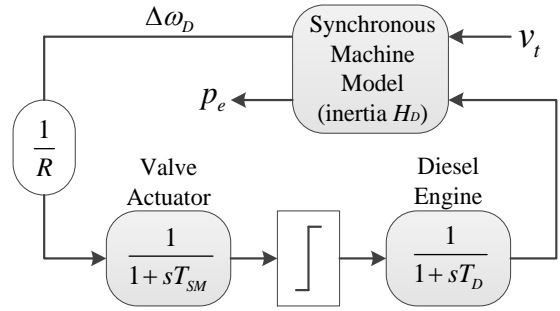

Fig. 2. Simplified block diagram of the diesel unit (DU) model

### 3.2 Battery storage

The block diagram of the battery storage unit is shown in Fig. 3. The battery represents an ideal voltage source with internal resistance $R_s$. A typical VSG control, as discussed in [35], is used for the DC/AC converter, as shown in Fig. 3.

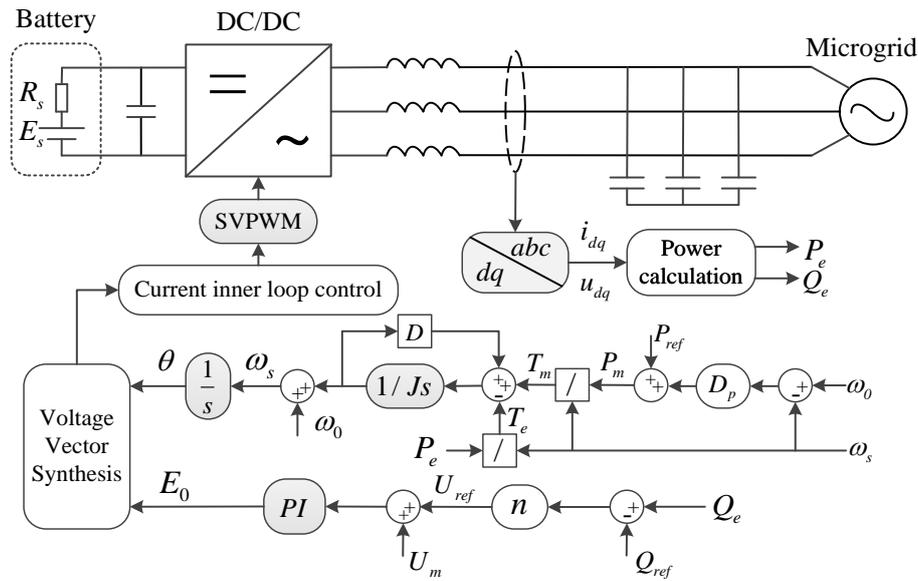

Fig. 3. Block diagram of battery storage unit

In fact, the swing equation is an important process of VSG,

$$J\frac{d\omega_s}{dt} = T_m - T_e - D(\omega_s - \omega_0) \quad (1)$$

Where $J$ is the moment of inertia (kg·m²); $T_m$, $T_e$, and $D$ are the virtual mechanical torque (N·m), virtual electromagnetic torque (N·m), and damping coefficient (N·m/rad), respectively; $D_p$ is droop coefficient; $\omega_s$ is the rotor angular frequency of the VSG; $\omega_0$ is the rated angular frequency of the VSG.

Owing to the existence of $J$, the grid-connected inverter has inertia in the power and frequency dynamic process; and due to the equivalent damping $D$, the inverter-type grid-connected power generation device also has the ability to dampen the grid power oscillation. These two variables are of great significance for the improvement of microgrid operational performance.

A $\omega$-$P$ droop is used to emulate the governor of VSG control, which can be represented as

$$P_m = P_{ref} + D_p(\omega_0 - \omega_s) \quad (2)$$

Hence, the battery storage automatically adjusts the active power output and provides inertia and primary FR support for the system.

### 3.3 PV System

A two-stage PV generation is used in this study, and its block diagram is illustrated in Fig. 4.

The PV array is modeled by a single-diode equivalent circuit [36]. A 3-phase DC/AC inverter connected to a microgrid bus, along with a DC/DC boost converter connected at the DC side.

In the typical PV generation control, as shown in Fig. 4, the DC/AC inverter regulates the DC-link voltage to its reference value. The DC/DC converter controls the voltage of the PV array to track the maximum power operating point. $V_{MPPT}$ is compared with $V_{PV}$ to generate the duty cycle of the boost converter through the PI controller. The maximum power can be achieved by adjusting the duty ratio of the boost circuit. For the typical control of PV generation, PV generation works in the MPPT mode and does not participate in the FR of the microgrid system.

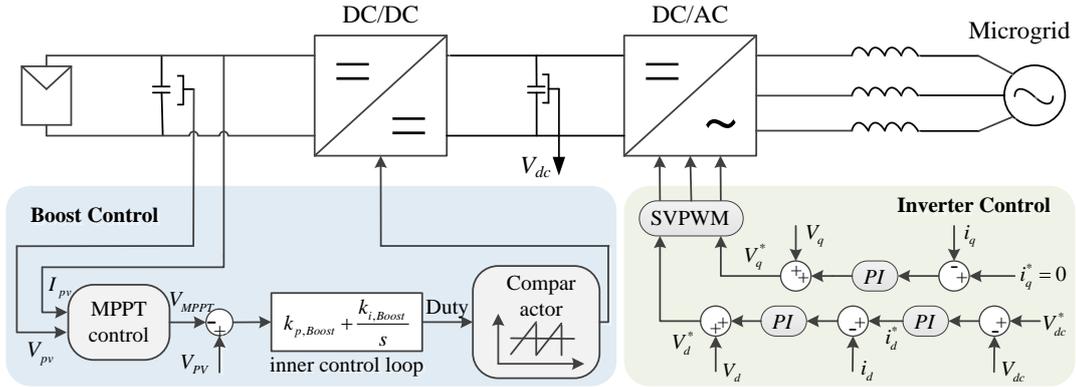

Fig. 4. Block diagram of the two-stage PV generation model

DU and the battery storage unit perform FR, while PV generation operates in the MPPT mode. Additionally, a DU of large capacity and a battery storage unit are essential to maintain the frequency within the allowance range. If a PV system can provide frequency support, it can improve the FR ability of the microgrid and reduce the required capacity of DU and energy storage. In fact, some papers have discussed the frequency support strategy for PV systems. As mentioned in the introduction, there are a few shortcomings to these strategies. VSG seems to be an admirable method for DGs providing frequency support. However, the typical VSG control is not adapted for two-stage PV generation.

(1) As seen in Fig. 3, a constant DC voltage source is required for a typical VSG. Thus, the active power control loop of the DC/AC inverter can be utilized to realize the emulated swing equation. However, for two-stage PV generation, the active power control loop is used to regulate the DC-link voltage.

(2) The PV system has no inherent inertia or available reserve power. Its output power is random and is determined by the environmental conditions. However, for a typical VSG, as seen in Fig. 3, DGs have a known reserve power.

Therefore, this study focuses on designing an improved VSG strategy for two-stage PV generation.

## 4. The proposed virtual synchronization control of PV generation

### 4.1. PV array model

The single diode equivalent model is a widely used engineering model for PV arrays, and its accuracy suffices for the purpose of this study. As given below:

$$I_{pv} = I_{ph} - I_0(e^{\frac{V_{pv}+R_s I_{pv}}{V_T}} - 1) - \frac{V_{pv} + R_s I_{pv}}{R_{sh}} \tag{3}$$

$$V_T = AkT/q \tag{4}$$

$$I_{ph} = \frac{S}{S_{STC}} I_{sc,STC}[1 + \alpha(T - T_{STC})] \tag{5}$$

$$I_0 = (I_{ph} - \frac{V_{oc}}{R_p})(e^{\frac{qV_{oc}}{AkT}} - 1)^{-1} \tag{6}$$

$$V_{oc} = V_{oc,STC}[1 + \beta(T - T_{STC})] + V_{T0}(\frac{S}{S_{STC}}) \tag{7}$$

where $I_{pv}$ is the PV output current; $I_{ph}$ is the photogenic current; $I_0$ is the diode reverse saturation current; $q$ is the electron charge, $q=1.6 \times 10^{-19}$ C; $V_{pv}$ is the voltage; $R_s$ is the series resistance; $R_{sh}$ is the parallel resistance; $A$ is the ideal diode factor; $S$ is the irradiance; $V_{oc}$ and $I_{sc}$ are the PV array open-circuit voltage and short-circuit current, respectively; $S_{STC}$ = 1000 W/m² and $T_{STC}$ = 25 ℃ are the standard test conditions; $I_{sc, STC}$ and $V_{oc, STC}$ are short-circuited current and open-circuit voltage under standard test conditions, respectively; and $α$ and $β$ are the thermal correlation coefficients.

In this study, SunPower's SPR-305E-WHT-D (number of series $N_s$ = 5, number of parallels $N_p$ = 66) PV array was selected. The values of the parameters are listed in Table A2 of Appendix.

### 4.2. Power reserve control for the PV system

Unlike WTG, PV generation does not have rotor kinetic energy. Therefore, in order to participate in FR, it is necessary to reserve a part of the active power for PV generation without energy storage. The authors have proposed a power reserve control method in an early study [37] and are introduced here.

The power reserve ratio $d$ is defined as

$$d = \frac{P_{mpp} - P_{de}}{P_{mpp}} \tag{8}$$

Where $P_{mpp}$ is the maximum power of the PV, $P_{de}$ is the de-loaded power of the PV, and $1-d$ is the power utilization ratio in this study.

A power reserve control for a PV system can be realized through a de-loaded power-voltage curve. Using the data of the PV array model in this study, the de-loaded power-voltage curve with $d = 20\%$ can be represented as a piecewise linear curve as in (9), as shown in Fig. 5. The fitting parameters are given in Table A3 of Appendix.

$$P_{de}(V_{pv}) = \begin{cases} a_1 V_{pv}, & 0 \leq V_{pv} < V_{p1} \\ a_2 V_{pv} + b_2, & V_{p1} \leq V_{pv} < V_{p2} \\ a_3 V_{pv} + b_3, & V_{p2} \leq V_{pv} < V_{p3} \\ a_4 V_{pv} + b_4, & V_{p3} \leq V_{pv} \end{cases} \quad (9)$$

A power tracking control rule was designed as in (10),

$$\begin{aligned} \triangle P_{pv}(i) &= P_{pv}(i) - P_{de}(V_{pv}(i)) \\ V_{ref}(i+1) &= V_{refold}(i) + sign(\triangle P_{pv}(i))dV_{pv}(i) \end{aligned} \quad (10)$$

The power reserve control procedure is illustrated in Fig. 5. Taking the de-loaded power-voltage curve as the boundary, the P-V characteristic curve is divided into the left and right segments. Assuming the initial operating point is $A(V_A, P_A)$ point located in the left area of the PV characteristic curve. Clearly, $P_A$ is greater than $P_{de}(V_A)$ located at the de-loaded power-voltage curve. Thus, according to the control rule of equation (9), $\Delta P_{pv}$ is positive and $V_{ref}(i+1)$ will increase. That is, the initial operating point "A" moves towards the intersection point $B(V_{de}, P_{de})$. As the operating point stays on the left area, $\Delta P_{pv}$ is always positive and the PV voltage continues to increase until it reaches the intersection point $B(V_{de}, P_{de})$.

Similarly, if PV initially stays on the right area of the P-V characteristic curve, $\Delta P_{pv}$ is always negative and $V_{ref}(i+1)$ continually decreases toward the target operating point $B(P_{de}, V_{de})$ through the control rule of equation (10). Note that this is automatically a power tracking process regardless of the irradiance conditions.

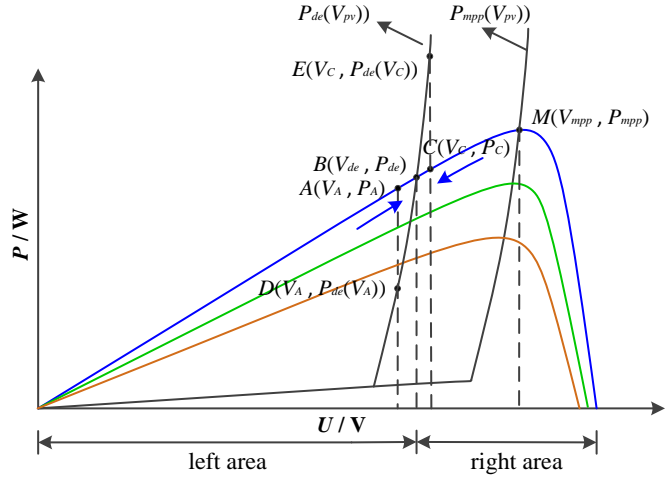

Fig. 5. Principle of de-loaded power point tracking

To prevent the oscillation of the operating point around the target point, a variable-step tracing method can be used, as in (11).

$$dV_{pv} = \begin{cases} \dfrac{k_1 \triangle P_{pv}}{V_{pv}^3} & |\triangle V| < \triangle V_{max} \\ sign(\triangle P_{pv})\triangle V_{max} & |\triangle V| > \triangle V_{max} \end{cases} \quad (11)$$

Where, $\Delta V_{max}$ is the maximum voltage step. In this study, $\Delta V_{max} = 0.5$ V.

### 4.3 The principle of the proposed VSG strategy

As well-known the power-angle characteristic equation in power system, the output power of the synchronous generator can be expressed,

$$P_s = \frac{EU}{X} \sin \delta \tag{12}$$

$$\delta = \int (\omega_s - \omega_g) dt \tag{13}$$

Where $P_s$ is output active power of the synchronous generator, $E$ is the amplitude of the synchronous generator electromotive force, $U$ is the amplitude of the grid voltage vector, $\delta$ is the power angle between two vectors, $X$ is the virtual impedance, $\omega_s$ is the rotational speed of the synchronous generator electromotive force, $\omega_g$ is the rotational speed of grid-tied voltage.

The output active power versus power angle curve is depicted in Fig. 6. While, the typical PV array P-V characteristic curve is given in Fig. 7.

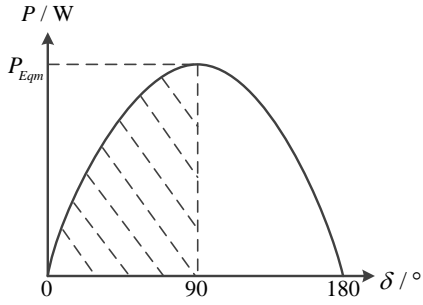 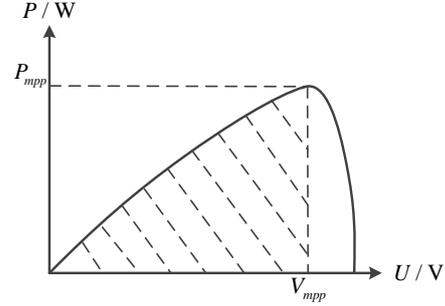

Fig. 6. Power angle characteristic curve        Fig. 7. PV array P-V characteristic curve

Obviously, the two curves have some similarities. In Fig. 6, during $\delta=0$ to 90°, $P_s$ increases with $\delta$, and the rate of change decreases. There is an extreme point at $\delta=90°$. In Fig. 7, in $V_{pv}=0$ to $V_{mpp}$ area, $P_{pv}$ increases with $V_{pv}$, and the rate of change decreases. An extreme point is located at $V_{pv} = V_{mpp}$.

Although, the shape of two curves still is not the same. However, based these intrinsic similarities, we consider that $V_{pv}$ can be analogized to the power angle $\delta$, which renders PV array voltage directly varies with grid frequency variation, and generate a similar "natural" inertial behavior that output power automatically increase or decrease when grid frequency variation.

Considering initial steady operating state, $\delta_0$ is assumed as the initial the power angle in virtual synchronous generator. Hence,

$$\begin{aligned} \delta &= \delta_0 + \int \triangle \omega \, dt \\ \triangle \omega &= \omega_s - \omega_g \end{aligned} \tag{14}$$

Based on the principle that $V_{pv}$ analogies to the power angle $\delta$,

$$V_{pv} = V_{de} + \int dV dt \tag{15}$$

Where, $V_{de}$ is the initial voltage, $dV$ is the differential voltage.

In discrete control, (15) can be re-expressed as,

$$\begin{aligned} V_{pv} &= V_{de} + \sum \Delta V \\ \Delta V &= V_{pv}(k+1) - V_{pv}(k) \end{aligned} \tag{16}$$

Compared (14) and (16), the $\Delta V$ can analogy to $\Delta \omega$. Simply, we assume $\Delta \omega$ has a linear relationship with $\Delta V$, that is

$$\Delta V = A \Delta \omega \tag{17}$$

Where, $A$ is an analogic coefficient.

**4.4 The proposed VSG strategy**

Based on the above analysis, the proposed VSG for two stage PV system is given in Fig. 8. The virtual mechanical torque $T_m$ is from the droop control, namely:

$$T_m = \frac{P_{de} + \Delta P}{\omega_s} \tag{18}$$

And

$$\Delta P = D_p (\omega_0 - \omega_g) \tag{19}$$

The rotational speed of VSG $\omega_s$ subtracting the grid rotational speed $\omega_g$ obtains $\Delta \omega$. $\Delta \omega$ multiplies an analogic coefficient $A$ to obtain PV voltage step $\Delta V$.

The swing equation block is the same with the swing equation, as in (1).

The rotational speed of VSG $\omega_s$ subtracting the grid rotational speed $\omega_g$ obtains $\Delta \omega$. $\Delta \omega$ multiplies an analogic coefficient $A$ to obtain PV voltage step $\Delta V$.

There is a potential accumulation process in the power curve change process, as in

$$P_{de} = P_{de}(V_{pv} - \Delta V(1)) \tag{20}$$

Where, $V_{pv}$ is the current PV array voltage. In each control period, $V_{pv}$ will update a new value. That is

$$V_{pv}(k+1) = V_{pv}(k) - \Delta V = V_{pv}(0) - \sum \Delta V \tag{21}$$

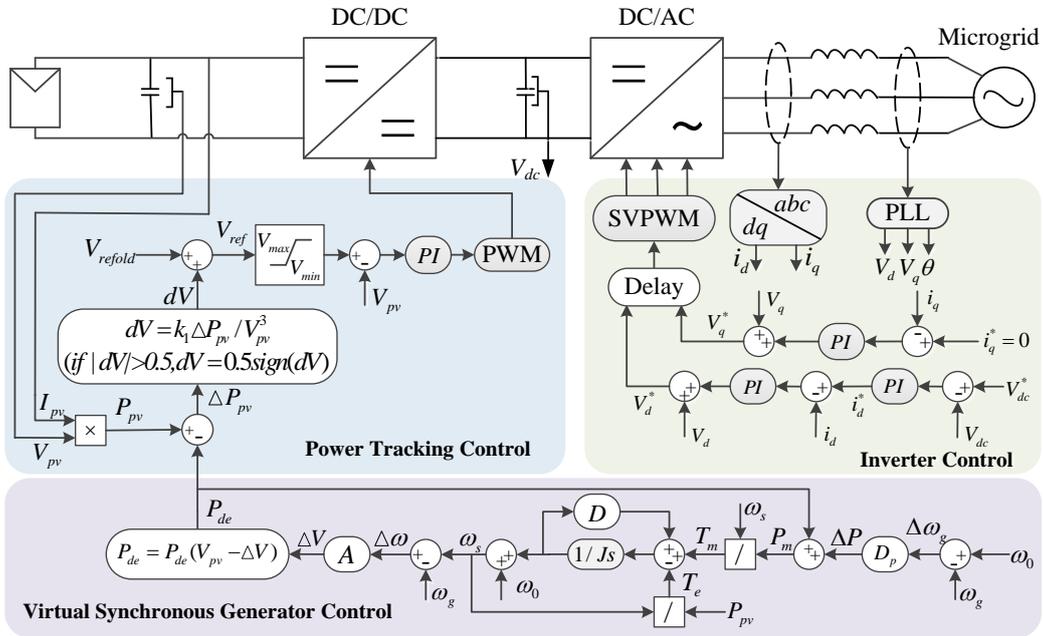

Fig. 8. Block diagram of the proposed PV array participating in FR strategy

Noted in the power curve tracking control, subtracting $\Delta V$ means the de-loaded curve moves right and the intersection moves up toward, which results to the increase of output power. The process can be illustrated in Fig. 9.

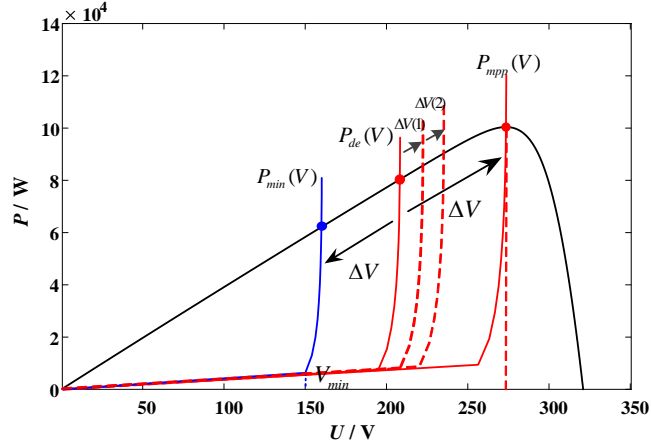

Fig. 9. The principle of power curve tracking control

$P_{de}(V_{de})$ is initial de-loaded power curve. $V_{de}$ subtracting $\Delta V(1)$ results in the de-loaded power curve moving right in the next control period. The power of intersection of the de-loaded power curve and P-V characteristic curve also increases. And in next-next control period, $V_{pv}$ subtracting $\Delta V(2)$, that is $V_{de}-\Delta V(1)-\Delta V(2)$, the de-loaded power curve continues to move up toward and the power of intersection increases. It is a potential accumulation process. The intersection is only equilibrium operating point for power curve tracking control.

When $\Delta \omega_g = 0$, the reference power $P_{de}$ from the de-loaded power curve directly feeds back to the virtual mechanical torque $T_m$. Thus, $\Delta \omega$ is zero and $\Delta V$ is zero. That is, under the condition of $\Delta \omega_g = 0$, the swing equation block is disabled and only the power reserve control is enabled.

When the grid frequency varies, $\Delta P$ is added to $P_{de}$. Affected by the block of the swing equation, $\Delta V$ is not equal to zero and gradually changes with $\Delta \omega_g$. $\Delta V$ is added to the measured PV array voltage $V_{pv}$. It shifts the de-loaded power-voltage curve. The deviation of the de-loaded power-voltage curve causes the target operating point to shift as the output power changes.

And a flow chart of the proposed VSG strategy is shown in Fig. 10.

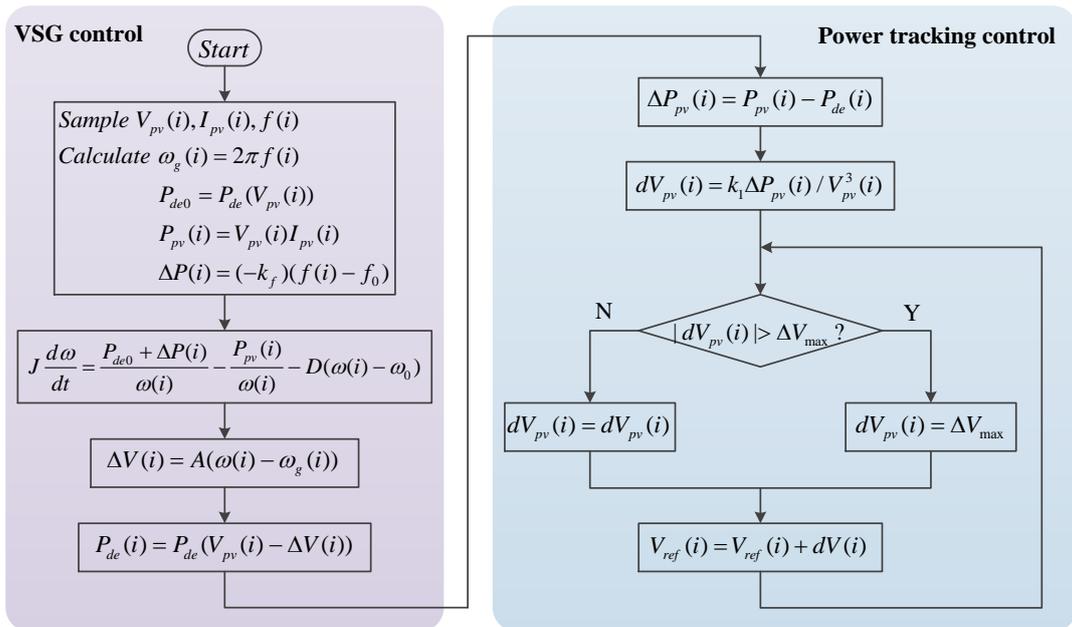

Fig. 10. The flow chart of the proposed VSG control flow chart

It should be noted that the proposed strategy requires the PV system to operate at the up-hill area of the P-V characteristic curve. This is because the output power of the PV system increases with the PV voltage at the up-hill segment, which is the correct direction for frequency support.

Otherwise, if PV operate at the downhill segment, the output power of the PV system rapidly decreases with the PV voltage, and the power change direction subsequently deteriorates the frequency control. To prevent the PV system from operating at the downhill segment, a saturation limiter is added to $V_{ref}$ in the control block diagram, as shown in Fig. 8.

The upper limit is set to $V_{mpp}$ to ensure that the PV system always operates at the up-hill segment. The lower limit is set to the minimum allowable voltage of d=40% de-loading curve (see in Fig.9). In this study, $V_{min}$ =150 V.

$V_{mpp}$ is an unknown value, as it is influenced by the environment and needs to be estimated. We have designed a simple method to rapidly estimate $V_{mpp}$, as described in the following section.

### 4.5. Maximum power voltage $V_{mpp}$ estimation method

In the up-hill segment of the P-V characteristic curve, $P_{pv}$ is almost linearly related to the PV voltage, and it can be approximately expressed as

$$P_{pv}(V_{pv}) = kV_{pv} \tag{22}$$

Where, $k$ is slope of the oblique line.

On the other hand, the maximum power-voltage curve can be approximated by a piecewise linear curve, and its expression is given.

$$P_{mpp}(V_{pv}) = \begin{cases} c_1 V_{pv}, & 0 \leq V_{pv} < V_{d1} \\ c_2 V_{pv} + d_2, & V_{d1} \leq V_{pv} < V_{d2} \\ c_3 V_{pv} + d_3, & V_{d2} \leq V_{pv} < V_{d3} \\ c_4 V_{pv} + d_4, & V_{d3} \leq V_{pv} \end{cases} \tag{23}$$

where $c_1 － c_4$ and $d_2 － d_4$ are the fitting coefficients, which are given in Table A2.

As shown in Fig. 11, $A_1(V_{pv1}, P_1)$, $A_2(V_{pv2}, P_2)$, and $A_3(V_{pv3}, P_3)$ are the points on the P-V characteristic curve of irradiance 200 W/m², 600 W/m², and 1000 W/m², respectively.

Oblique lines are made across $A_1$, $A_2$, and $A_3$. Clearly, there is an intersection between the oblique line corresponding to equation (22) and the maximum power-voltage curve of equation (23). Their intersections are at points $C_1$, $C_2$, and $C_3$, respectively. $D_1$, $D_2$, and $D_3$ are the actual maximum power points (MPP).

Using PV array modeling in this study, the voltage deviation between the points $C_1$ and $D_1$ is 0.4069 V, the voltage deviation between $C_2$ and $D_2$ is 0.1653 V, and the voltage deviation between $C_3$ and $D_3$ is 0.0058 V. This error is small and can be considered negligible.

The main reason for this is that the slope of the maximum power-voltage curve is rather large. Therefore, the error between the intersection point and the actual MPP is very small. The greater the slope of the maximum power curve is, the smaller the error is. Thus, the voltage at the intersection point can be approximated as the MPP voltage.

The PV output power during the start-up segment (segment ① in Fig. 11) is very small and can be ignored here. Then, by solving equations (22) and (23), the approximate MPP voltage can be calculated from:

$$V_{mpp} = \begin{cases} -d_2/(c_2 - k), V_{d1} \leq V_{mpp} < V_{d2} \\ -d_3/(c_3 - k), V_{d2} \leq V_{mpp} < V_{d3} \\ -d_4/(c_4 - k), V_{d3} \leq V_{mpp} \end{cases} \tag{24}$$

Further, $k$ can be obtained from the current measurement of the PV power and voltage, as in

$$k = \frac{P(V_{pv})}{V_{pv}} \tag{25}$$

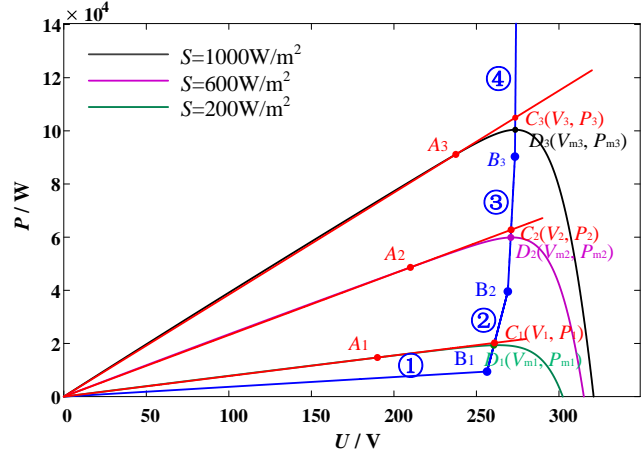

(a) Overall picture

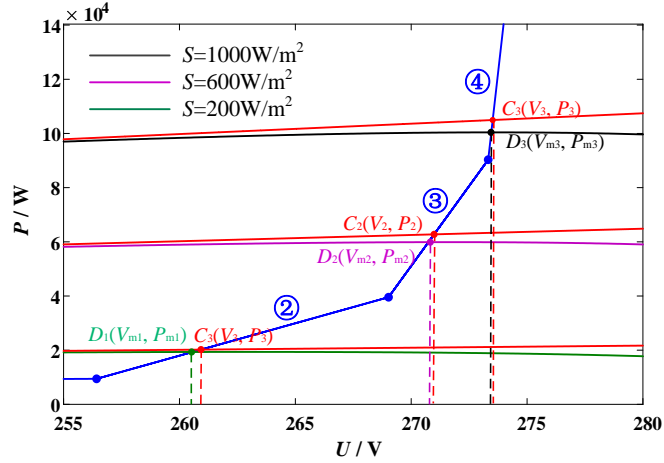

(b) Partial picture

Fig. 11. Estimation method of maximum power voltage $V_{mpp}$

## 5. Time-domain simulation results

In order to verify the effectiveness of the proposed VSG control strategy, an island PV-diesel-storage microgrid modeling, as described in Fig. 1, was built using MATLAB/SIMULINK 2018 Student Suite Version, MathWorks, Natick, MA, USA. The values of the parameters in the island MG are presented in Table 2. The values of the parameters of two-stage PV generation are listed in Table A4 of Appendix.

### 5.1 Comparison with the existing VSG method

[29] proposed a PRC-VSG control for de-loaded PV system participating in FR. The similar auxiliary frequency control loop also can be found in paper [30,38], and paper [39-41] for wind turbines. The control diagram of the PRG-VSG is depicted in Fig.12.

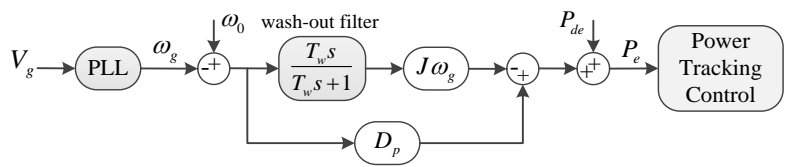

Fig.12. the control diagram of PRC-VSG

The control diagram can be expressed in (26)

$$P_f = -J\omega_g \frac{d\omega_g}{dt} + D_p(\omega_0 - \omega_g) \qquad (26)$$

Where $\omega_g$ is from the PLL and the $d\omega_g/dt$ is obtained through a wash-out filter with 0.1 time constant, as seen in Fig. 12.

PLL brings some measurement delay and error. Further, the differential value $d\omega_g/dt$ is sensitive to noise. They may reduce the control stability and deteriorate the FR performance. In addition, for the PRC-VSG control, the inertial support is generated through the power tracking control. It is totally different from the inertial response of the SG which is naturally generated depending on the angular frequency variation.

By comparison, the inertia response of our proposed strategy is depended on the mimic swing equation blocks, which does not require the $d\omega_g/dt$ information. Even more, the affection of $\omega_g$ from PLL is weakened because the rotational speed of the SG electromotive force $\omega_s$ is self-generated. The power angle $\delta$ is replaced with PV voltage $V_{pv}$, which renders PV array voltage directly response to grid frequency variation and generate a "natural" inertial behavior that output power automatically increase or decrease. It can better mimic the inertia behavior of SG.

The advantages of the proposed strategy can be summarized in Table.3.

**Table 3. The advantage of the proposed strategy**

| Index | Advantages |
|---|---|
| 1 | Adaptive. It applied to a DC-DC converter, while the grid-side inverter can still be used to control DC-link voltage. |
| 2 | Simple. No complex PV array modeling or estimation algorithm required. |
| 3 | Economic. No energy storage required, and no additional irradiance or temperature sensors required. |
| 4 | Robust. "Naturally" generates the inertial response through directly regulating PV array voltage, and does not require $d\omega_g/dt$ measurement. |

In simulations, the proposed VSG strategy is compared with the PRC-VSG control and that PV does not participate in FR. The droop coefficient $D_p$ and the inertia $J$ are identical for the PRC-VSG control and the proposed VSG strategy. In details, $D_p$ is selected as 40 and the $J$ is selected as 1.06. Three groups of simulations were carried out to evaluate the performance of the frequency support of two-stage PV generation.

**5.2 Sudden change in load under standard conditions**

In this group, the environment of the PV system is the standard condition ($S = 1000$ W/m$^2$, $T = 25$ ℃). In Case 1, the load is suddenly increased, whereas the load is decreased in case 2.

**Case 1: Step increase in load**

The PV generation initially maintains at a 20% power reserve, and the load increases by 10 kW at 30 s, resulting in a frequency dip. The simulation results are given in Fig. 13.

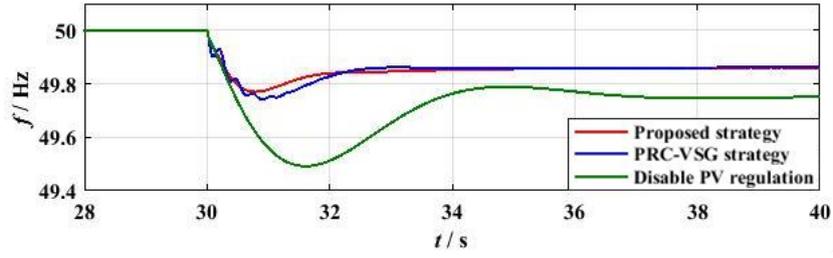

(a) System frequency

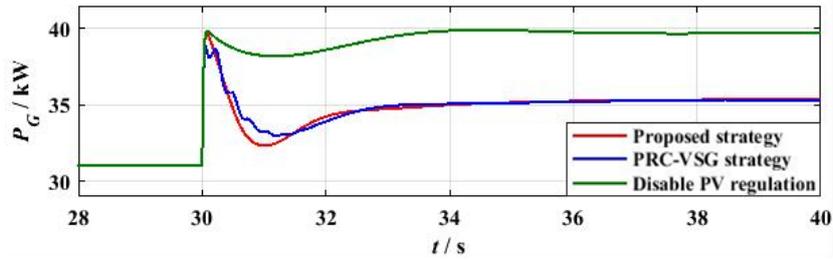

(b) Active power output of the DU

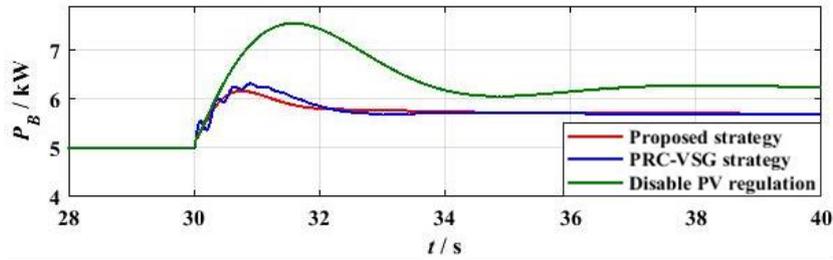

(c) Active power output of battery storage.

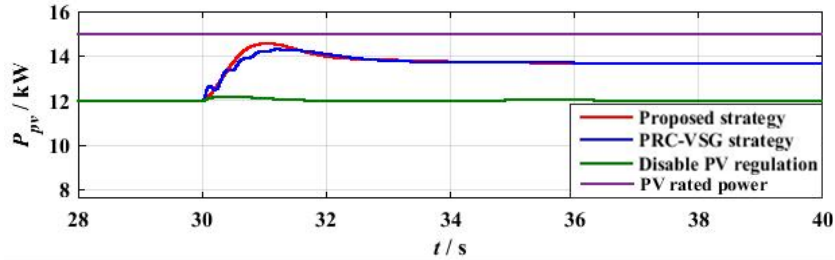

(d) Active power output of the PV system

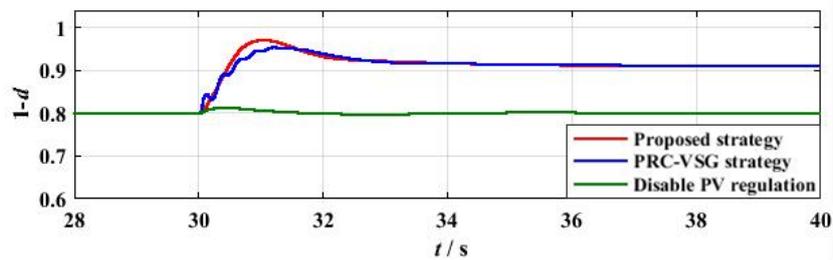

(e) Power utilization ratio

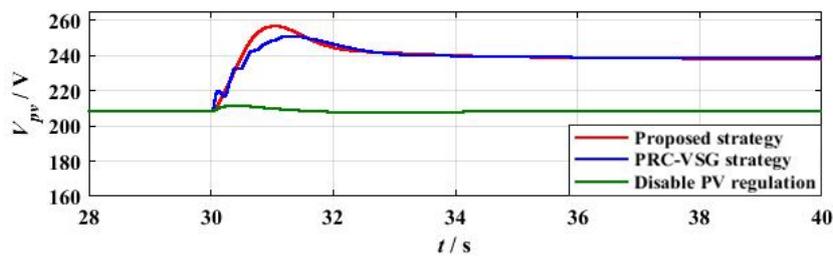

(f) PV voltage

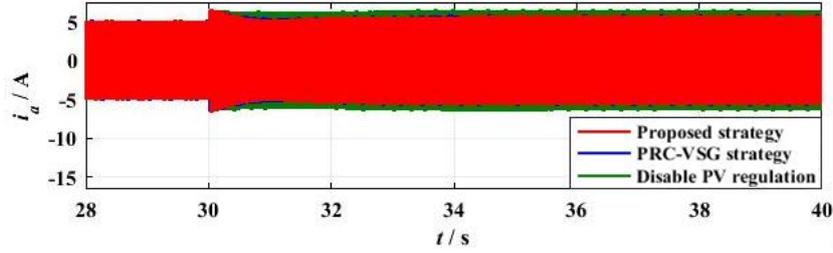

(g) System current

Fig. 13. System dynamic responses when load suddenly increases

Although three 15 kW PV generators are located in the microgrid at different positions, the environment is identical in this study, and the output power of three PV generations is almost the same. Therefore, only one active power curve of the PV generation is given in this case.

As shown in Fig. 13, before 40 s, the total output power of PV generation was $12 \times 3$ kW, which tracks the given 20% of the power reserve. That is, when $\Delta f = 0$, the PV system operates in the PRC mode.

After 30 s, the system frequency decreases. For PVs without VSG, its output power is maintained at 12 kW. The steady frequency of the microgrid is 49.75 Hz, and the FN is 49.49 Hz. FR depends on the DU and the battery storage unit. The output power of the DU and the battery storage unit increase with frequency dips. The peak power of the DU is 39.91 kW, and the steady power is 39.75 kW. The peak power of the battery storage unit is 7.55 kW and the steady power is 6.25 kW.

For PVs with PRC-VSG, as the system frequency decreases, the PV system generates more active power for frequency support. The peak power of each PV generation is 14.3 kW, and their steady power is 13.66 kW. The peak power of the DU is 38.82 kW, and the steady power is 35.32 kW. The peak power of the battery storage unit is 6.32 kW and the steady power is 5.7 kW. The steady frequency of the microgrid is 49.86 Hz, 0.11 Hz higher than the one without PV frequency. And FN is 49.74 Hz, 0.25 Hz higher than the one without PV FR. However, some power and frequency oscillations occur in the transient process.

Comparatively, for PVs with the proposed VSG, the steady state is the same as that of the PRC-VSG strategy due to the same droop control loop. However, their transient processes will be different due to different inertial responses. In the proposed VSG, the peak power of each PV generation, DU and battery storage unit is 14.56 kW, 39.68 kW and 6.16 kW respectively. The FN of the proposed VSG is 49.77 Hz, 0.03 Hz higher than PRC-VSG. Significantly, the power and frequency show a smooth dynamic process. This also validates the discussion in the previous section.

The inertia response for PRC-VSG is depended on $d\omega_g/dt$. $\omega_g$ from PLL input into a wash-out filter to obtain $d\omega_g/dt$ (see in Fig. 12).

As we all know, $d\omega_g/dt$ measurement is affected by the filter delay and attenuation, which further affect the control performance. The $d\omega_g/dt$ curve of sudden load increase of 10 kW in 30 s is given in Fig.14, which shows an intensive oscillation.

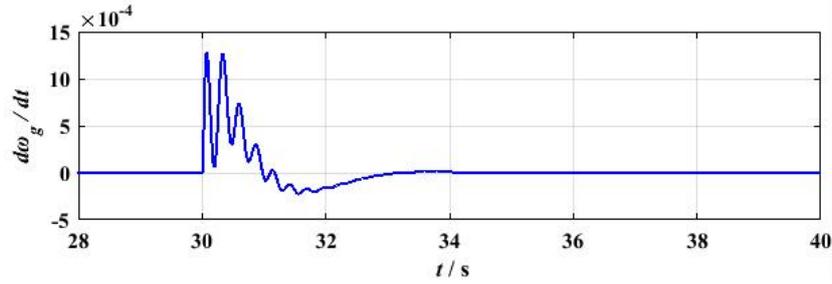

Fig. 14. Simulation result of d$\omega_g$/dt in PRG-VSG control

The simulation results show that in the proposed and PRC-VSG strategy, PV generators can share the frequency regulation task with DU and the battery storage and improve the frequency response of the system. However, the FR effect of the proposed VSG is better than that of PRC-VSG.

**Case 2: Step decrease in load**

In the case of PV generations with 20% power reserve at the initial stage, the load is suddenly decreased by 10 kW at 30 s, resulting in a frequency rise event. The simulation results for case 2 are depicted in Fig. 15.

For the PV without VSG, the output power of each PV generation is always 12 kW. The frequency peak is 50.5 Hz. The steady frequency is 50.25 Hz. The output power of the DU and the battery storage system decreases as the frequency increases. The lowest power of DU is 22.01 kW, and the steady power is 22.25 kW. The lowest power of the battery storage system is 2.47 kW and the steady power is 3.75 kW.

For PVs with PRC-VSG, the active power of PV generations decreases as the system frequency increases. The lowest power of each PV generation is 9.7 kW, and its steady power is 10.34 kW. The lowest power of DU is 23.05 kW, and the steady power is 26.68 kW. The lowest power of the battery storage system is 3.7 kW and the steady power is 4.3 kW. The highest frequency is 50.26 Hz, 0.24 Hz lower than the one without VSG. Finally, the new steady state frequency is 50.14 Hz, 0.11 Hz better than the one without VSG.

Comparatively, for PV generations with the proposed VSG, they reduce more active power to inhibit the increase of system frequency in the transient process. The lowest power of each PV generation, DU and battery storage unit is 9.45 kW, 22.19 kW and 3.86 kW respectively. The steady state is the same as that of the PRC-VSG strategy. However, the highest frequency of the proposed VSG is 50.23 Hz, 0.03 Hz lower than PRC-VSG.

From the simulation results of the two cases, through the proposed VSG, the PV system can better and more smoothly co-regulate microgrid frequency with DU and the battery storage unit. It improves both the frequency peak (or nadir) and steady frequency.

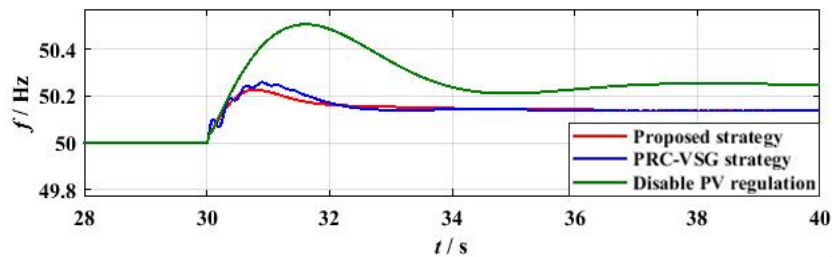

(a) System frequency

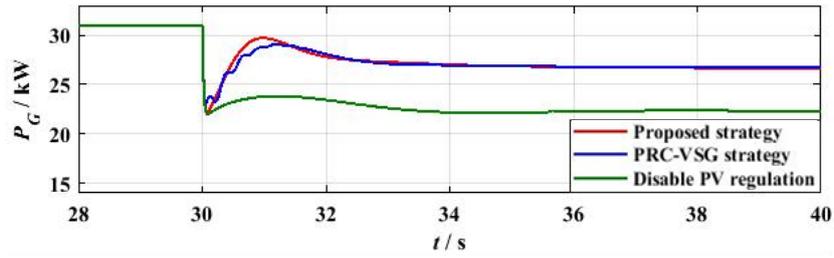
(b) Active power output of the DU

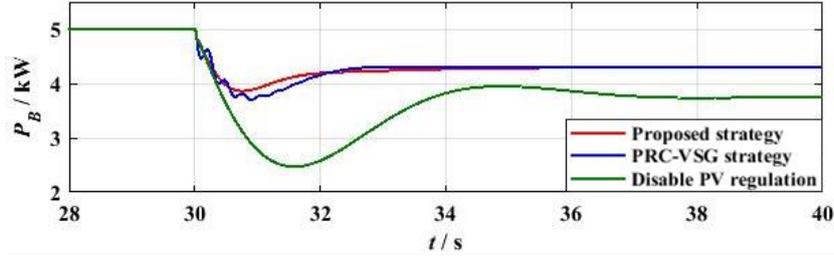
(c) Active power output of battery storage

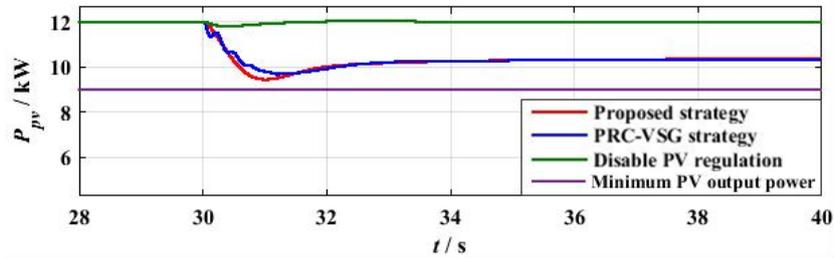
(d) Active power output of the PV system

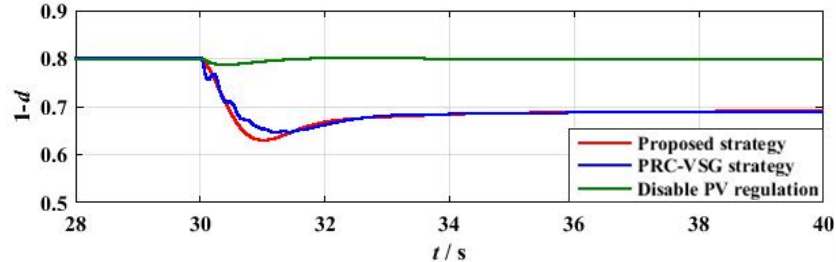
(e) Power utilization ratio

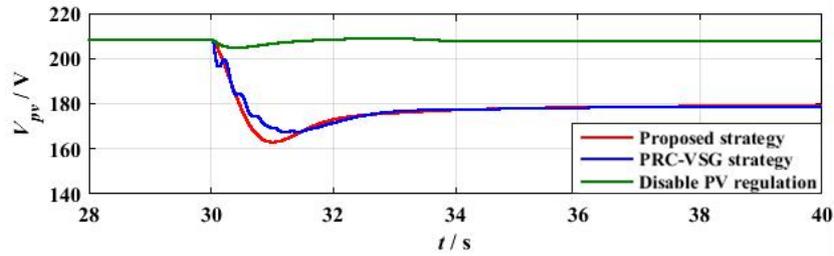
(f) PV voltage

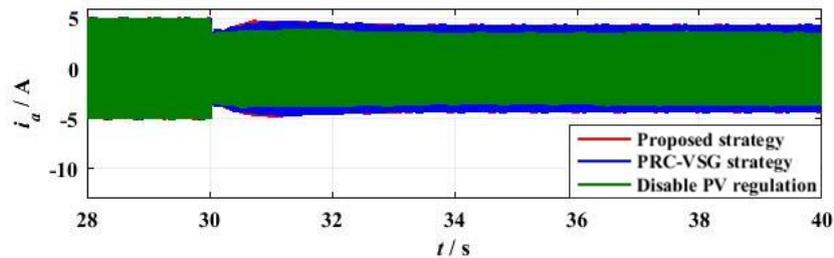
(g) System current

Fig. 15. System dynamic responses when load suddenly decreases

## 5.3 Load step change under change in irradiance

This group tests the proposed VSG strategy considering the change in irradiance. In case 3, the load step increases, while the irradiance for the PV system continually decreases. In case 4, the load suddenly decreases when the irradiance continues to increase. The change in direction of the irradiance is opposite to that of the load. In case 5, the load randomly changes under the actual condition of irradiance measurement.

**Case 3: Irradiance decreases and load step increases:**

The PV system also maintains at a 20% power reserve at initial, and the irradiance $S$ continuously decreases with a slope of 5 W/m$^2$/s from the $t = 30$ s (as shown in Fig. 16). At $t = 45$ s, the load suddenly increases by 8 kW. The simulation results are shown in Fig. 17.

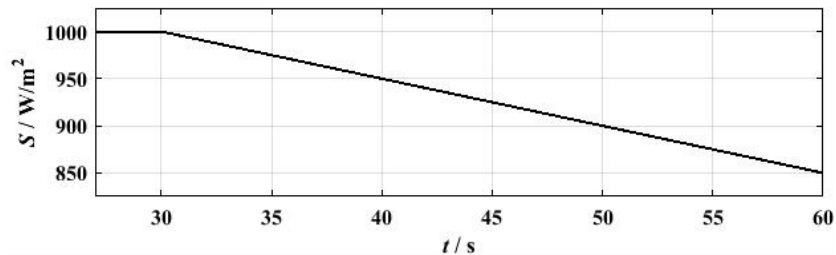

Fig. 16. Given irradiance

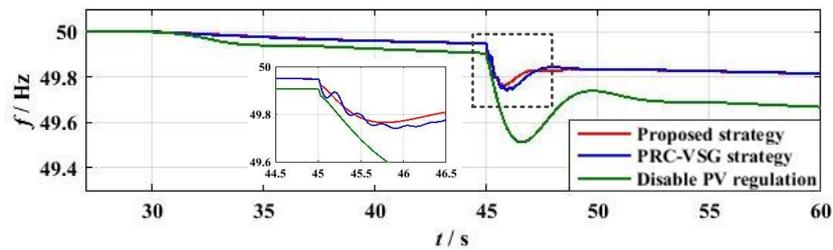

(a) System frequency

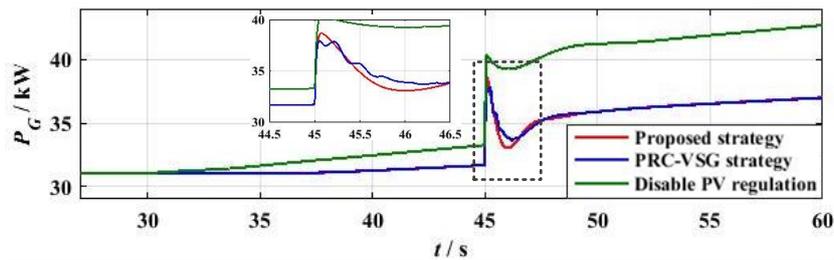

(b) Active power output of the DU

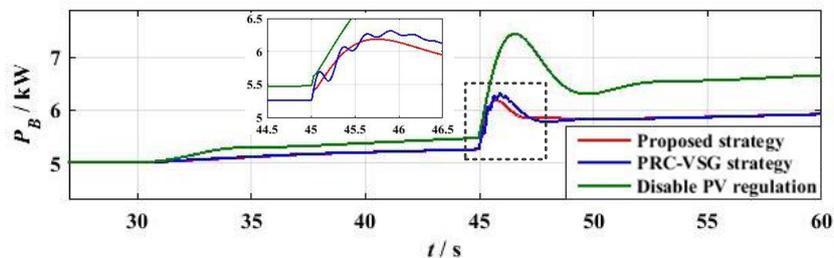

(c) Active power output of battery storage

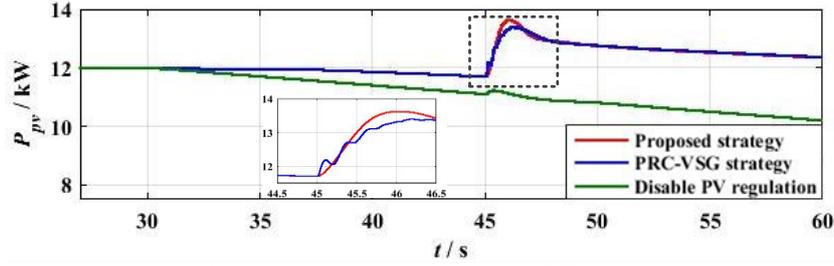

(d) Active power output of the PV system

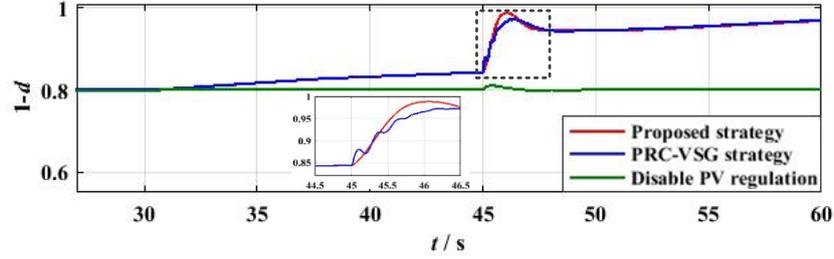

(e) Power utilization ratio

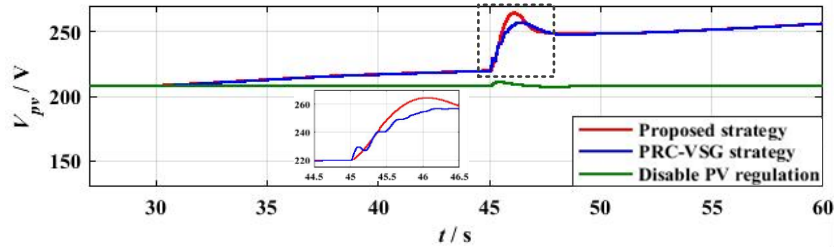

(f) PV voltage

Fig. 17. System dynamic responses under decrease in irradiance

As shown in Fig. 17, during 30–45 s, with the irradiance decreasing, the output power of the PV system decreases, and the system frequency shows a slight decrease. After 45 s, owing to the sudden increase in load, the system frequency significantly decreases.

For PVs without VSG, the PV output power continues to decrease as the irradiance decreases, while the power reserve ratio remains at 20%, as shown in Fig. 18 (e). It is validated that PRC control in this study can track the pre-set power reserve ratio under varying irradiance conditions. DU and the battery storage unit perform FR. The FN for PV without VSG is 49.51 Hz. Moreover, affected by the decrease in irradiance, the steady frequency continuously decreases slightly. At 60 s, the system frequency is 49.67 Hz.

For PVs with the proposed and PRC-VSG, at 30–45 s, affected by slight decrease in frequency caused by the irradiance, the output power is slightly greater than that of the disabled VSG PV system, as shown in Fig. 17 (d). Then, the PV system output power changes depending on the variation of the frequency and irradiance. In fact, the two strategies add additional active power to the initial de-loaded power. They show a potential of adaption with variation in the irradiance. And the final frequency of the two strategies at 60s is the same. However, FN for PV with PRC-VSG strategy is 49.74 Hz, 0.23 Hz higher than that of the PV without VSG. Additionally, at 60 s, the system frequency is 49.81 Hz. While, FN for PVs with the proposed VSG is 49.76 Hz, which is 0.02 Hz higher than that of PRC-VSG strategy.

**Case 4: Irradiance increases and load step decreases:**

The irradiance S continuously increases with a slope of 5 W/m$^2$/s from the simulation time $t$ = 30 s, as shown in Fig. 18. When $t$ = 45 s, the load suddenly decreases by 8 kW. The simulation results are shown in Fig. 19.

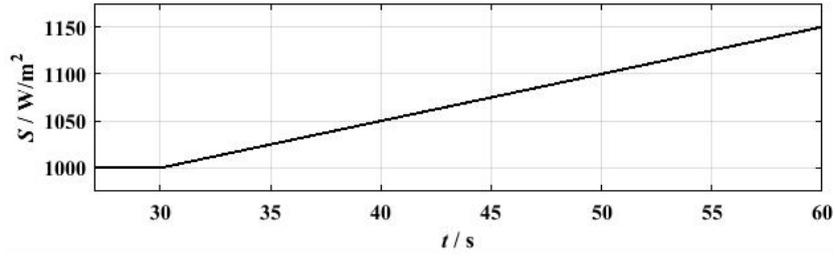

Fig.18. Given irradiance

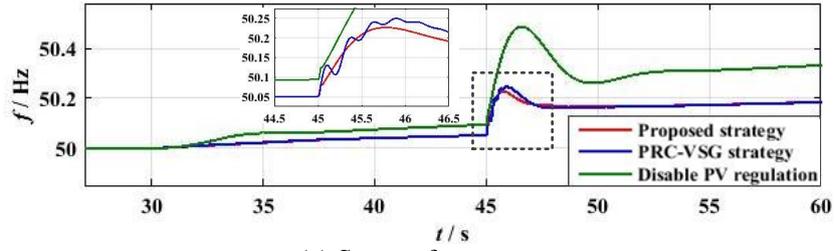

(a) System frequency

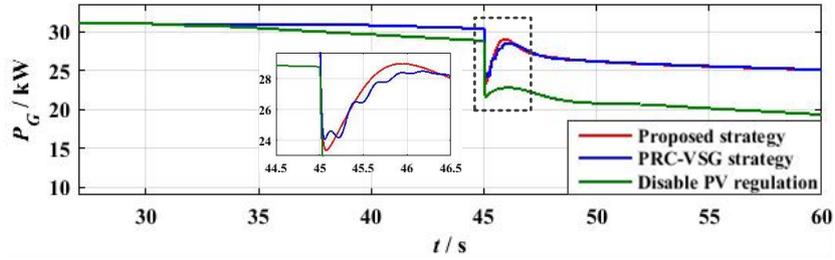

(b) Active power output of the DU

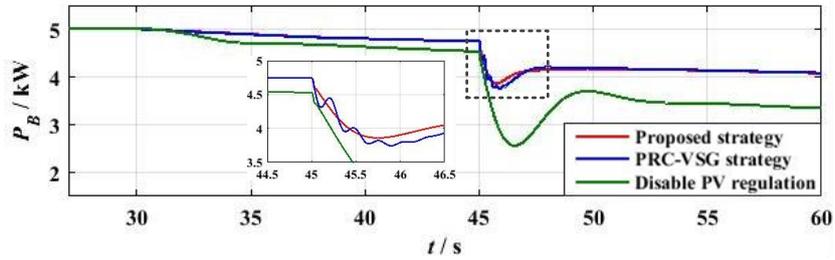

(c) Active power output of battery storage

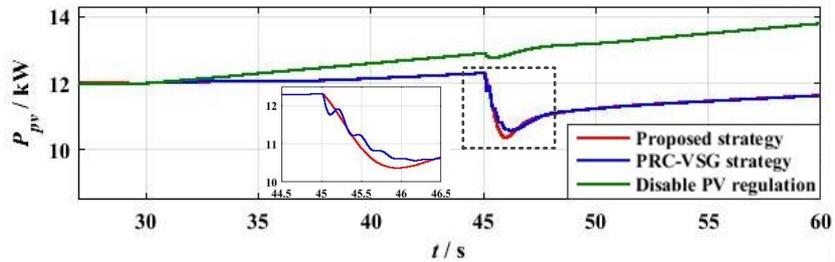

(d) Active power output of the PV system

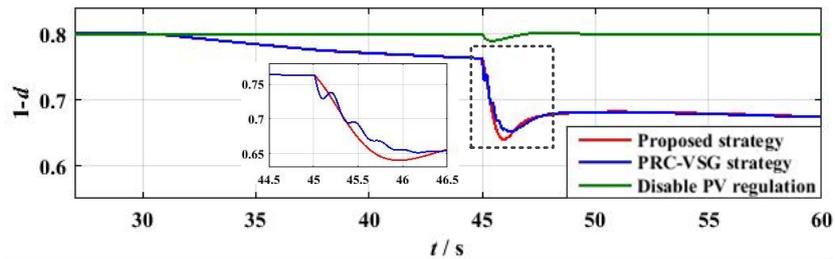

(e) Power utilization ratio

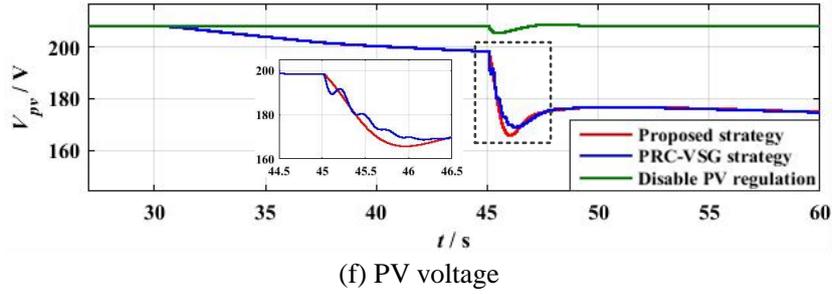

(f) PV voltage

Fig. 19. System dynamic responses under irradiance increases conditions

As shown in Fig. 19, during 30–45 s, the irradiance increases cause the output power of the PV system to increase and the frequency slightly increases. At 45 s, the load sudden decrease results in a frequency rise event.

Disabled VSG PV systems do not participate in FR. Its frequency peak is 50.49 Hz. The PV output power continues to increase as the irradiance increases, while the power reserve ratio remains at 20%. As the irradiance increases, the system frequency continues to increase, and the steady frequency rises to 50.33 Hz at 60 s.

For PVs with PRC-VSG, its frequency peak is 50.25 Hz, 0.24 Hz lower than that of the PV without VSG. Additionally, at 60 s, the system frequency is 50.18 Hz.

For PVs system with the proposed VSG, its frequency peak is 50.23 Hz, 0.02 Hz lower than the one with PRC-VSG. At 60 s, the system frequency also rises to 50.18 Hz, which is the same as that of PRC-VSG strategy. The PV system reduces a part of the active power on the de-loaded power according to the frequency variation.

**Case 5: Random change in irradiance and load:**

The real irradiance and the load data imported in the simulation modeling, which last 160 s and are shown in Fig. 20. The simulation results are shown in Fig. 21.

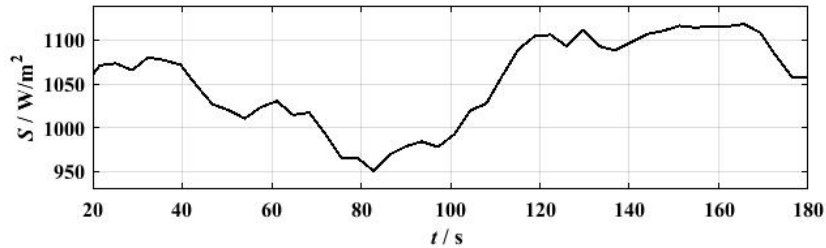

(a) Irradiance change

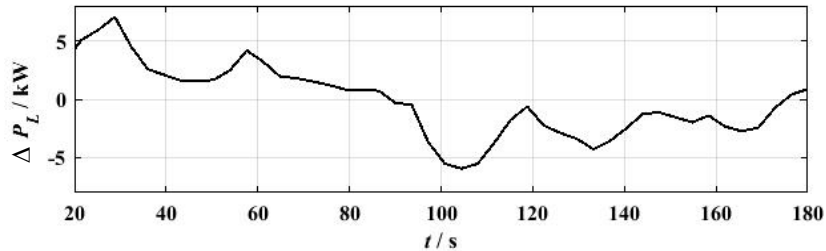

(b) Load change

Fig. 20. Actual irradiance and load curve

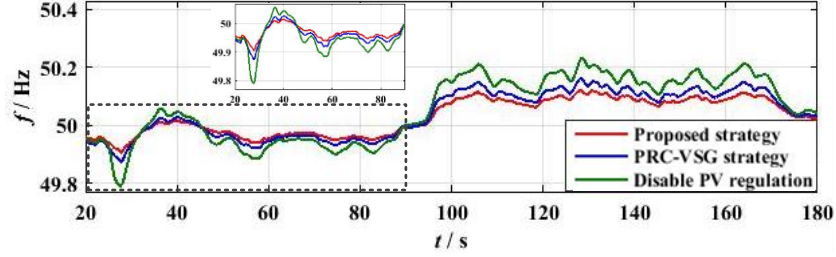
(a) System frequency

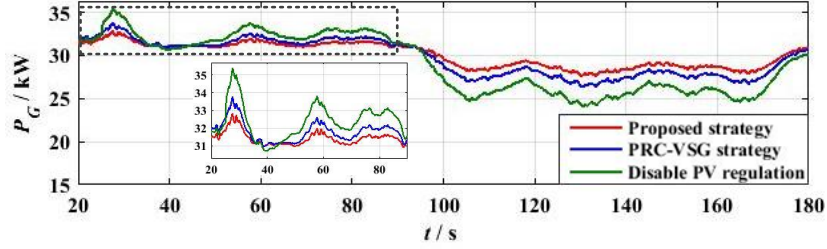
(b) Active power output of the DU

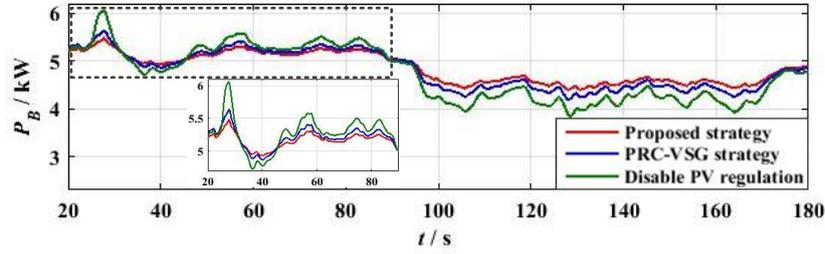
(c) Active power output of battery storage.

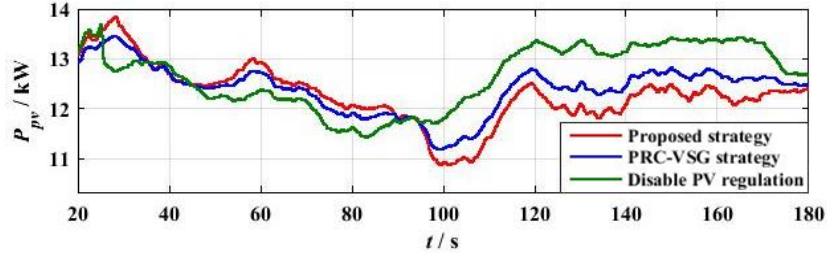
(d) Active power output of the PV system

Fig. 21. System dynamic responses when irradiance and load change randomly

As shown in Fig. 21, the frequency fluctuation for the PV system with VSG is considerably smaller than that for the PV system without VSG.

In microgrids including PV generations without VSG, the maximum frequency deviation is 0.23 Hz, the root mean square is 50.0567 Hz, and the root mean square error is 0.1247 Hz.

In PV generations with PRC-VSG strategy, the maximum frequency deviation is 0.16 Hz, the root mean square is 50.0392 Hz, and the root mean square error is 0.0886 Hz.

In contrast, when the PV system with the proposed VSG participates in the FR of the microgrid, the maximum frequency deviation is only 0.12 Hz, the root mean square is 50.0304 Hz, and the root mean square error is 0.0689 Hz. PVs with the proposed VSG significantly improve the frequency response performance of the microgrid under actual load and irradiance conditions.

**5.4 Effects of PV power penetration level**

PV power penetration level (PV-PPL) is defined as the ratio of PV power generation to load demand electricity.

The total capacity of the microgrid load was maintained at 72 kW in this study. 30%, 50%,

and 70%, three PV-PPLs are tested in this case. The battery storage is identical for all penetration levels. To achieve different penetration levels, the capacity of PV power generation increases, and the capacity of the DU decreases. The capacity of the DGs in the island microgrid is summarized in Table 4.

Under standard environment conditions, the load suddenly increases by 10 kW at 40 s, and the comparative simulation results of different penetration levels are shown in Fig. 22 and Fig. 23.

Table 4. Capacity of DGs in island microgrid

| DG | Penetration 30% | Penetration 50% | Penetration 70% |
| --- | --- | --- | --- |
| PV | $9 \times 3$ kW | $15 \times 3$ kW | $21 \times 3$ kW |
| DU | 45.4 kW | 31 kW | 16.6 kW |
| Battery storage | 5 kW | 5 kW | 5 kW |

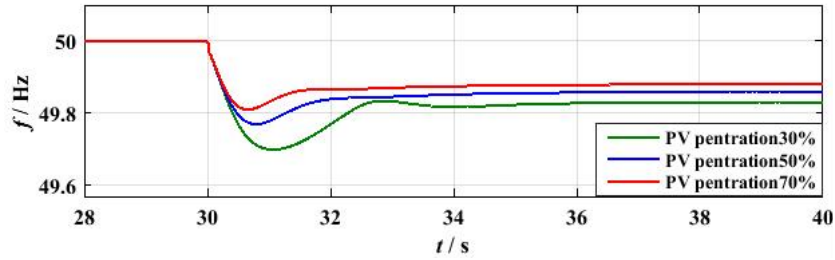

(a) System frequency with PV participating in FR.

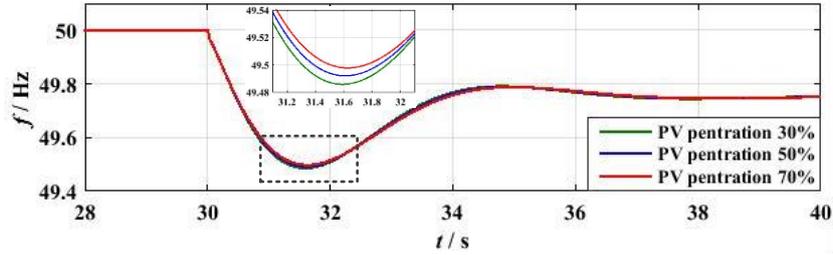

(b) System frequency with PV not participating in FR.

Fig. 22. System frequency at different PV penetration levels

In Fig. 22, if PV uses a disabled VSG strategy, the microgrid frequency is very similar for the different power penetrations. FN is 49.49 Hz, and the steady frequency is 49.75 Hz. When PV system does not participate in FR, the frequency response is determined by the DU and battery. The battery is identical for all three scenarios. The DU models for all three scenarios are identical except the given power reference. Therefore, frequency nadirs are almost same for all three scenarios.

Comparatively, when the PV uses the VSG strategy for FR, the dynamic response of the system frequency is quite different. In the case of 30% penetration level, FN is 49.7 Hz and the steady frequency is 49.83 Hz; in the case of 50% penetration level, FN is 49.77 Hz and the steady frequency is 49.86 Hz. In the case of 70% penetration level, FN is 49.81 Hz and the steady frequency is 49.88 Hz.

To compare the dynamic behavior of DU with PV participating in FR, the output active power of these DGs for the different penetration levels is depicted in Fig. 23.

With the increase in the penetration level of the PV generation, FN and the steady frequency is better. In our view, compared with DU, the PV system has a faster and better active power control ability.

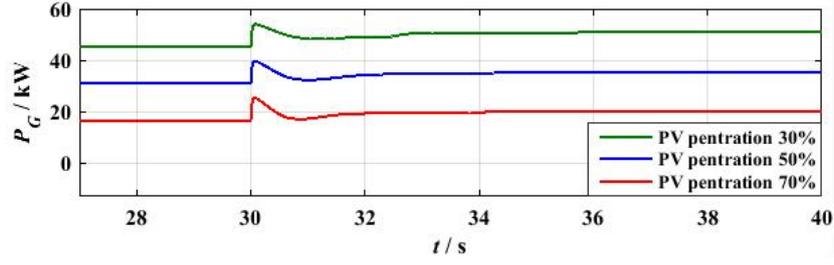
(a) Active power output of the DU

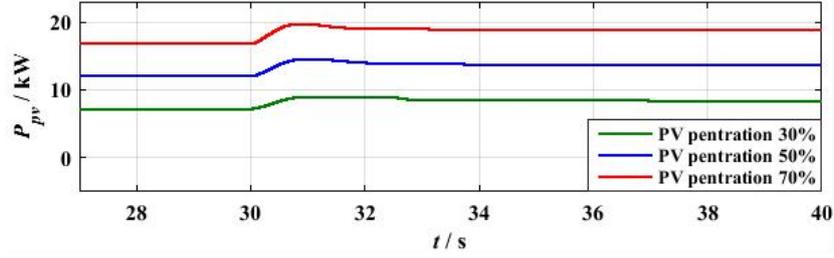
(b) Active power output of the PV system

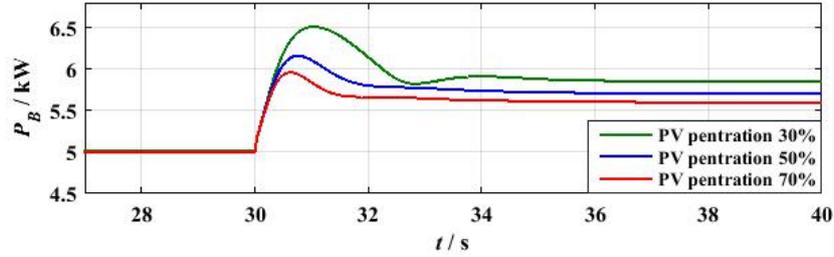
(c) Active power output of battery storage.

Fig. 23. Active power output at different PV penetration levels with PV participating in FR

## 6. Conclusion

This paper introduces a novel VSG control for two-stage PV generation to provide frequency support for island microgrids. Based on the similarities of the synchronous generator power-angle characteristic curve and the PV array P-V characteristic curve, PV voltage $V_{pv}$ can be analogized to the power angle $\delta$. Therefore, it applies to the DC/DC converter to mimic the swing equation and droop governor of SGs. A simple power reserve control based on the de-loaded power-voltage curve is designed to initially reserve a part of the power. Through the proposed VSG strategy, additional active power is generated and added to the de-loaded power to provide frequency support.

A real island microgrid modeling was selected as a study case, and the different PV-PLLs were investigated through simulation. The simulation results clearly indicate that the proposed VSG strategy can effectively make PV generation provide a frequency support under different irradiance conditions, which improves the FR ability of the microgrid. With an increase in the penetration level of the PV generation, the FN and the steady frequency show a better performance.

## Acknowledgments

This work was supported by National Natural Science Foundation of China [grant numbers U18866601, 2018~2022]; National Natural Science Foundation of China [grant numbers

52077030, 2020-2024]; National Natural Science Foundation of Jilin Province, China [grant number 20190201289JC, 2019~2021].

**Declaration of Competing Interest**

The authors declare that they have no known competing financial interests or personal relationships that could have appeared to influence the work reported in this paper.

**Appendix**

Table A1  Diesel generator parameters

| Parameter | Symbol | Value |
|---|---|---|
| Nominal power | $P_{DU}$ | 31kW |
| Droop | $R_{DU}$ | 5% |
| DU time constant | $T_D$ | 0.5s |
| Valve actuator servomechanism time constant | $T_{SM}$ | 0.05s |
| DU inertia time constant | $H_D$ | 3s |

Table A2  Key parameters of a single PV array

| Parameter | Symbol | Value |
|---|---|---|
| Open-circuit voltage under standard conditions | $V_{oc,STC}$ | 62.4V |
| Short-circuit current under standard conditions | $I_{sc,STC}$ | 5.96A |
| Short-circuit current thermal correlation coefficient | $\alpha$ | 0.061745 |
| Open-circuit voltage thermal correlation coefficient | $\beta$ | -0.27269 |
| Ideal diode factor | A | 0.94504 |
| Open circuit voltage | $V_{oc}$ | 64.2V |
| Short circuit current | $I_{sc}$ | 5.96A |
| Series resistance | $R_s$ | 0.37152Ω |
| Parallel resistance | $R_{sh}$ | 269.5934Ω |

Table A3  Fitting parameter values of equation (9) (17)

| Parameter | Value | Parameter | Value |
|---|---|---|---|
| $a_1$ | 38.6888 | $c_1$ | 36.77674654 |
| $a_2$ | 2459.477551 | $c_2$ | 2391.15873 |
| $a_3$ | 12688.5 | $c_3$ | 11803.25581 |
| $a_4$ | 120668 | $c_4$ | 75417.5 |
| $b_2$ | -472053.8024 | $d_2$ | -603717.4559 |
| $b_3$ | -2566957.6 | $d_3$ | -3135787.107 |
| $b_4$ | -25026693.6 | $d_4$ | -20523016.81 |
| $V_{p1}$ | 195 | $V_{d1}$ | 256.4229 |
| $V_{p2}$ | 204.8 | $V_{d2}$ | 269.0229 |
| $V_{p3}$ | 208 | $V_{d3}$ | 273.3229 |

Table A4　　Parameters of the PV generation

| Parameter | Symbol | Value |
|---|---|---|
| PV array MPP voltage | $V_{mpp}$ | 273.5V |
| PV array MPP current | $I_{mpp}$ | 368.28A |
| PV array series and parallel number | $n_p \times n_s$ | 66×5 |
| DC-link capacitance | $C_{dc}$ | 6mF |
| Grid inductance | $L_f$ | 250μH |
| Droop coefficient | $k_d$ | 50 |
| Virtual inertia coefficient | $k_i$ | 4 |

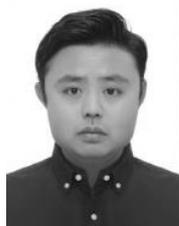
**CHENG ZHONG** (M'20) was born in JiangXi, China, in 1985. He received the B.S. degree from the Harbin Institute of Technology, Harbin, China, in 2007 and the M.S. and Ph.D. degrees in power systems and automation from China Agricultural University, Beijing, China, in 2010 and 2014, respectively.

He is currently an Assistant Professor with the School of Electrical and Electronic Engineering, Northeast Electric Power University, JiLin, China, where he has been since 2014. His current research interests include the area of power system distribute intelligent control, power system frequency regulation by renewable power.

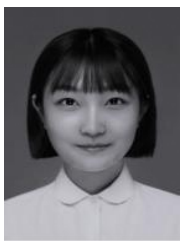
**HUAYI LI** (S'20) was born in Baoding, China, in 1998. She received the B.S. degree from Normal China Electric Power University Science and Technology College, in 2019. She is currently pursuing the M.E. degree in electrical engineering from Northeast Electric Power University. Her research interest includes active frequency control of high permeability photovoltaic island microgrid.

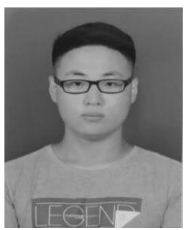
**YANG ZHOU** (S'20) was born in Qingdao, China, in 1996. He received the B.S. degree from Qufu Normal University, Rizhao, in 2018. He is currently pursuing the M.E. degree in electrical engineering from Northeast Electric Power University. His research interest includes power system frequency regulation by renewable power.

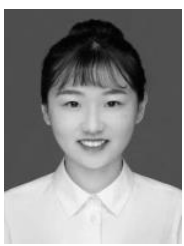
**YUEMING LV**(S'20) was born in Luoyang, China, in 1997. She received the B.S. degree from Northeast Electric Power University, in 2019. She is currently pursuing the M.E. degree in electrical engineering from Northeast Electric Power University. Her research interest includes power system frequency regulation.

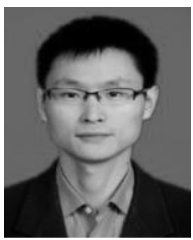
**JIKAI CHEN** (Member, IEEE) received the Ph.D. degree in electrical engineering from the Harbin Institute of Technology, Harbin, China; in 2011.He is currently an Associate Professor with the School of Electrical Engineering in Northeast Electric Power University. His current research interests include HVdc control technologies, analysis and control of power quality, and modeling analysis and control of power electronics dominated power systems.

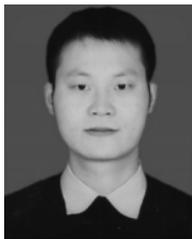 **YANG LI** (S'13 - M'14 - SM'18) was born in Nanyang, China. He received the Ph.D. degree in electrical engineering from North China Electric Power University (NCEPU), Beijing, China, in 2014.

He is currently an Associate Professor with the School of Electrical Engineering, Northeast Electric Power University, Jilin, China. From January 2017 to February 2019, he held a China Scholarship Council (CSC)-funded postdoctoral position at the Argonne National Laboratory, Lemont, USA. His research interests include power system stability and control, renewable energy integration, and smart grids. He also serves as an Associate Editor for IEEE ACCESS.